\documentclass[letterpaper, 12pt]{article}

\usepackage[utf8]{inputenc}
\usepackage[T1]{fontenc}
\usepackage{amsmath,amssymb,amsfonts}
\usepackage{algpseudocode}
\usepackage{algorithm}
\usepackage{graphicx}
\usepackage{textcomp}
\usepackage{booktabs}
\usepackage{array}
\usepackage{url}
\usepackage{notoccite}
\usepackage{amsthm}
\usepackage{bm}
\usepackage{cite}
\usepackage{color}
\usepackage{geometry}
\usepackage[colorlinks=true, citecolor=blue, filecolor=black, linkcolor=red, urlcolor=blue]{hyperref}

\title{\LARGE \bf
Deep Koopman-based Control of Quality Variation in Multistage Manufacturing Systems
\thanks{This paper addresses a typo found in the version to be published in the proceedings of the 2024 American Control Conference (ACC). Readers should pay attention to the formulation in Eq. \eqref{quality_control_formulation}. Specifically, Eq. \eqref{quality_control_formulation} aims to solve for a sequence of downstream commands that compensate for quality variations. However, the command terms $\Delta_X$ are not properly shown in the ACC version. Despite this typo, the results and conclusions presented in both versions of the paper remain unchanged.}
}

\author{Zhiyi Chen\thanks{Department of Mechanical Engineering, University of Michigan, Ann Arbor, MI, USA
        {\tt\small \{chzhiyi, xhuan, junni\}@umich.edu} }
        \and 
        Harshal Maske\thanks{Ford Motor Company, Dearborn, MI, USA
        {\tt\small \{hmaske1, deveshu, huanyis\}@ford.com} }
        \and
        Devesh Upadhyay\footnotemark[3]
        \and
        Huanyi Shui\footnotemark[3]
        \and
        Xun Huan\footnotemark[2]
        \and
        Jun Ni\footnotemark[2] \thanks{Global Institute of Future Technology, Shanghai Jiao Tong University, Shanghai, China}
}

\date{}

\begin{document}

\maketitle

\begin{abstract}

This paper presents a modeling-control synthesis to address the quality control challenges in multistage manufacturing systems (MMSs). 
A new feedforward control scheme is developed to minimize the quality variations caused by process disturbances in MMSs. Notably, the control framework leverages a stochastic deep Koopman (SDK) model to capture the quality propagation mechanism in the MMSs, highlighted by its ability to transform the nonlinear propagation dynamics into a linear one. 
Two roll-to-roll case studies are presented to validate the proposed method and demonstrate its effectiveness. The overall method is suitable for nonlinear MMSs and does not require extensive expert knowledge.

\end{abstract}

\section{Introduction}
\label{sec: Introduction}
Production processes in modern industry are often multistage manufacturing systems (MMSs), where the product undergoes multiple operation stages to reach its final form. Maintaining consistent quality in MMSs is challenging due to the intrinsic interdependence of operations among stages. Minor upstream deviations can propagate and aggregate, potentially leading to large downstream quality disruptions. Therefore, it is crucial for operators to detect and address abnormal quality variations at the earliest opportunity.

Conventional methods for quality variation reduction mainly rely on robust process design and statistical process control techniques \cite{shi2023process}. However, these methods are often limited by their inability to provide real-time responses to online disturbances. In contrast, the concept of in-process quality improvement (IPQI) focuses on online process monitoring and active process control, presenting a more effective approach to mitigating quality variations\cite{shi2023process}.

IPQI methods that can handle MMSs remain scarce~\cite{dreyfus2022virtual}. One prominent example is the Stream-of-Variation (SoV), widely recognized for its capability to model and analyze quality propagation in MMSs. SoV uses a linear state-space representation to capture the inter-stage propagation of variations, allowing one to predict quality variations stage-by-stage and develop control strategies to compensate for the predicted variations. For instance, SoV has been utilized to investigate dimensional error propagation and to develop a stochastic control law~\cite{djurdjanovic2007online}. This approach is extended in \cite{djurdjanovic2017multistage} to also incorporate uncertainty in noise characteristics. Moreover, an active control is designed in \cite{abellan2012quality} for variation reduction in computer numerical controlled (CNC) machining centers. Despite these successes, SoV is limited to machining and assembly applications due to its inherent linear representation. New methods are thus needed for quality modeling and control in nonlinear MMSs \cite{lee2022stream}.

Machine learning (ML) techniques have recently emerged as powerful tools to help fill this gap. Deep learning-based quality prediction for MMSs \cite{yan2021deep, zhang2021path, zhang2022contrastive, wang2023production} have been introduced to combine feature extraction and model learning into a unified framework, demonstrating good prediction accuracy for nonlinear processes. These methods all focus on modeling and prediction, and they have yet to consider any controller design or control application. Elsewhere, \cite{peres2019multistage} presented an ML classifier to detect defective products for quality control purposes, but it requires human expert intervention for system enhancement decisions. A fusion of a long-short term memory (LSTM) network and genetic algorithm is used in \cite{zhao2023novel} for final-stage quality prediction and process optimization, but the method cannot respond to quality disruptions on the intermediate products, necessitating delays to any corrective decisions. Overall, while the above examples offer broader applicability to nonlinear MMSs, they currently do not possess the ability for real-time quality control.

The key innovation and contribution of this paper is to develop a feedforward control scheme for active, in-situ compensation of quality variations. In particular, this new control algorithm makes use of a previously developed stochastic deep Koopman (SDK) model \cite{chen2023stochastic} that is capable of capturing the per-stage quality propagation within nonlinear MMSs. The effectiveness of the overall SDK-control framework, in terms of prediction accuracy and control performance, is demonstrated through two novel case studies of roll-to-roll (R2R) manufacturing processes.


\section{Methodology}
\label{sec: methodology}
Consider a production system of $N\geq 2$ stages. 
At the $k^{\rm th}$ stage (denoted by $S_k$), process measurements are represented by ${X_k=[x_{k,1}, \ldots, x_{k,p_k}]^\top}$ where $x_{k,i}$, ${i=1, 2, \ldots, p_k}$, is the $i^{\rm th}$ individual measurement from a total of $p_k$.
Similarly, quality indices are $Y_k=[y_{k,1}, \ldots, y_{k,q_k}]^\top$ where $y_{k,i}$, $i=1, 2, \ldots, q_k$, is the $i^{\rm th}$ quality index from a total of $q_k$.
The quality control task aims to minimize the variations in $Y_k$.
In Sec. \ref{sec: SDK}, we first summarize the SDK model adopted from~\cite{chen2023stochastic} for quality propagation, and then in Sec. \ref{sec: SDK_control}, we present the novel feedforward quality variation control.

\subsection{Modeling of Quality Propagation}
\label{sec: SDK}
\begin{figure}[ht]
\centerline{\includegraphics[width=\columnwidth]{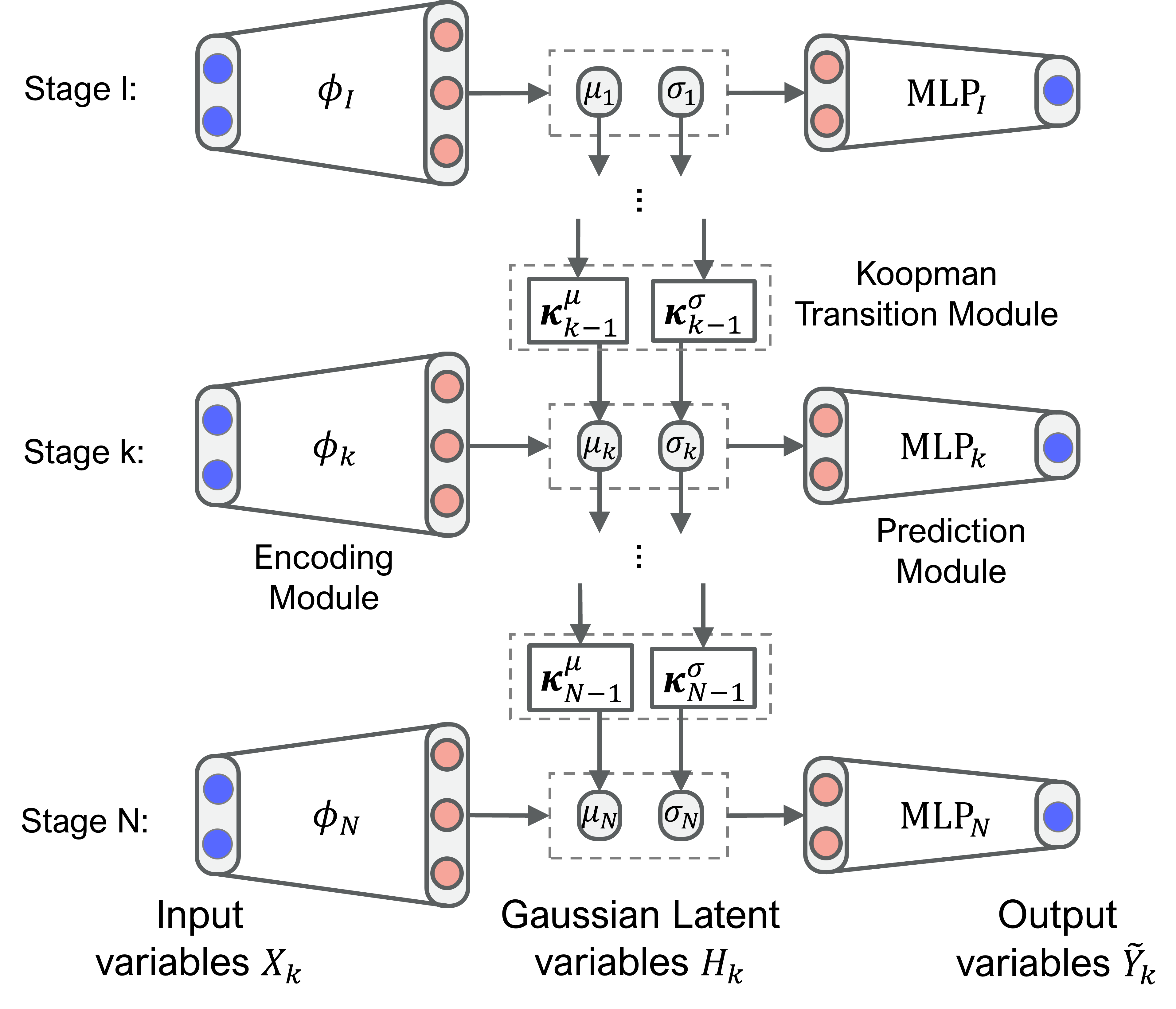}}
\caption{Modeling of quality propagation using an SDK framework. This approach yields a latent space with higher dimension compared to the original input space, due to the use of Koopman operators.}
\label{fig:fig1}
\end{figure}

 In an MMS, product quality at $S_k$ is causally affected by $X_1$ through $X_k$. Therefore, process-related information must be propagated from the upstream to enable downstream quality predictions. Using an encoding-decoding mechanism, we perform this propagation in a transformed space through a latent variable called quality indicator $H_k$. The transformation helps mitigate redundant information in $X_k$ and facilitates better modeling accuracy. More importantly, it allows us to seek a heuristic representation of the propagation mechanism in the Hilbert space. The proposed framework is depicted in Fig. \ref{fig:fig1} and described below.

At stage ${S_k \ (k\geq2)}$, the encoding module $\phi_k$ takes the local process measurements $X_k$ and transforms it into a temporal quality indicator $\hat{H}_k$. This transformation is achieved via a variational autoencoder (VAE) \cite{kingma2013auto}, which follows:
\begin{align} \label{VAE_description}
    \hat{H}_k &= \phi_k({X}_k) & \text{(Encoder)}\\
    \hat{H}_k  &\sim \mathcal{N}(\hat{\mu}_k, \hat{\sigma}_k^2) & \text{(Gaussian latent variable)}
    \\
    \tilde{X}_k &= \psi_k(\hat{H}_k) & \text{(Decoder)}
\end{align}
where $\hat{\mu}_k$ is the temporal mean, $\hat{\sigma}_k^2$ is the vector of temporal variance, and $\tilde{X}_k$ is the VAE reconstruction estimate of $X_k$ via the decoder $\psi_k$. Note that $\psi_k$ is absent in Fig. \ref{fig:fig1} since it is not used for prediction, but serves as a VAE reconstruction regularization term during the training process. $\hat{H}_k$ is then combined with the quality information propagated by a Koopman transition module $\mathcal{K}_{k-1}$ from the upstream to compute the quality indicator $H_k$.

Generally, the propagation of quality information in MMSs is nonlinear. We approximate this nonlinear propagation using linear embeddings discovered by Koopman operators \cite{lusch2018deep}. Koopman's theory posits that nonlinear dynamics have equivalent linear representations in the infinite-dimensional Hilbert space. In our case, an infinite-dimensional Koopman operator $\mathcal{K}_{k-1}$ is approximated by a finite-dimensional Koopman matrix $\mathbf{K}_{k-1}$, which facilitates a more tractable analysis of the nonlinear quality propagation in MMSs \cite{chen2023stochastic}.

Following this approximation, the dimension of $H_k$ is typically greater than that of $X_k$, and so the VAE's role is not dimension reduction but to find an appropriate Koopman transformation. Using the Koopman transition matrices, the quality indicators can be propagated linearly as:
\begin{equation} \label{Koopman_propagation}
    H_k = \hat{H}_k + \mathbf{K}_{k-1}H_{k-1}.
\end{equation}
Notably, our stochastic framework models $H_k\sim \mathcal{N}(\mu_{k}, \sigma^2_{k})$ and $H_{k-1} \sim \mathcal{N}(\mu_{k-1}, \sigma^2_{k-1})$ as independent Gaussian variables with mean $\mu$ and variance $\sigma^2$. This allows for a robust distributional description of the latent information. This distributional propagation is approximated by the transition of $\mu_k$ and $\sigma_k$ individually \cite{balakrishnan2021stochastic}:
\begin{align} 
\mu_k&=\hat{\mu}_k+\mathbf{K}^{\mu}_{k-1}\mu_{k-1} \label{KoopmanMean} 
\\
\ln{\sigma_k}&=\ln{\hat{\sigma}_k}+\mathbf{K}^{\sigma}_{k-1}\ln{\sigma_{k-1}} \label{KoopmanVariance}
\end{align}
where $\mathbf{K}^{\mu}_{k-1}$ and $\mathbf{K}^{\sigma}_{k-1}$ are Koopman operators for the mean and log standard deviation, respectively. Note that the Koopman propagation is applied to $\ln(\sigma)$ in \eqref{KoopmanVariance} to improve the numerical stability during training. Once the $\mu_k$ and $\sigma_k$ are obtained, $H_k$ can be sampled from the estimated distribution through reparameterization $H_k=\mu_k+\epsilon_k\sigma_k$, where $\epsilon_k \sim \mathcal{N}(0,1)$ is a standard Gaussian. Finally, $H_k$ can be utilized to predict quality indices $\tilde{Y}_k$ through a multilayer perceptron (MLP):
\begin{equation}
    \tilde{Y}_k = \text{MLP}_k(H_k).
\end{equation}

Upon assembling all modules, the overall SDK framework for MMSs can be trained end-to-end. One can also perform a two-step training described in \cite{chen2023stochastic}. In either case, the total loss function is
\begin{equation}
    \mathcal{L}_{\text{total}} = \sum_{i=1}^{N} \big( \rho_i \mathcal{L}_{\text{pred},i} + \theta_i \mathcal{L}_{\text{recon},i} + \omega_i \mathcal{L}_{\text{KLD},i} \big) \label{VAEtotalloss}
\end{equation}
where $\rho_i, \ \theta_i, \ \omega_i$ are weighting factors, and
\begin{align}
    \mathcal{L}_{\text{pred},k} &= \frac{1}{n} \sum^{n}_{i=1}\| (Y_k - \tilde{Y}_k)_i \|_2^2 \label{predloss}
    \\
    \mathcal{L}_{\text{recon},k} &= \frac{1}{n} \sum^{n}_{i=1}\| (X_k - \tilde{X}_k)_i \|^2_2 \label{reconloss}
    \\
    \mathcal{L}_{\text{KLD},k} &= D_{\text{KL}}\left(p(\hat{H}_k \mid X_k)\,\, \|\,\, \mathcal{N}(0,I)\right) \label{KLD}
\end{align}
with $n$ being the number of training points. Here the primary objective is to minimize the prediction loss $\mathcal{L}_{\text{pred}}$. Concurrently, the reconstruction loss $\mathcal{L}_{\text{recon}}$ acts as a regularization term to guide feature extraction, and the Kullback-Leibler divergence $\mathcal{L}_{\text{KLD}}$ further regularizes the Gaussian distributions in the latent space.

\subsection{Control of Quality Variation}
\label{sec: SDK_control}
\begin{figure}[ht]
\centerline{\includegraphics[width=0.7\columnwidth]{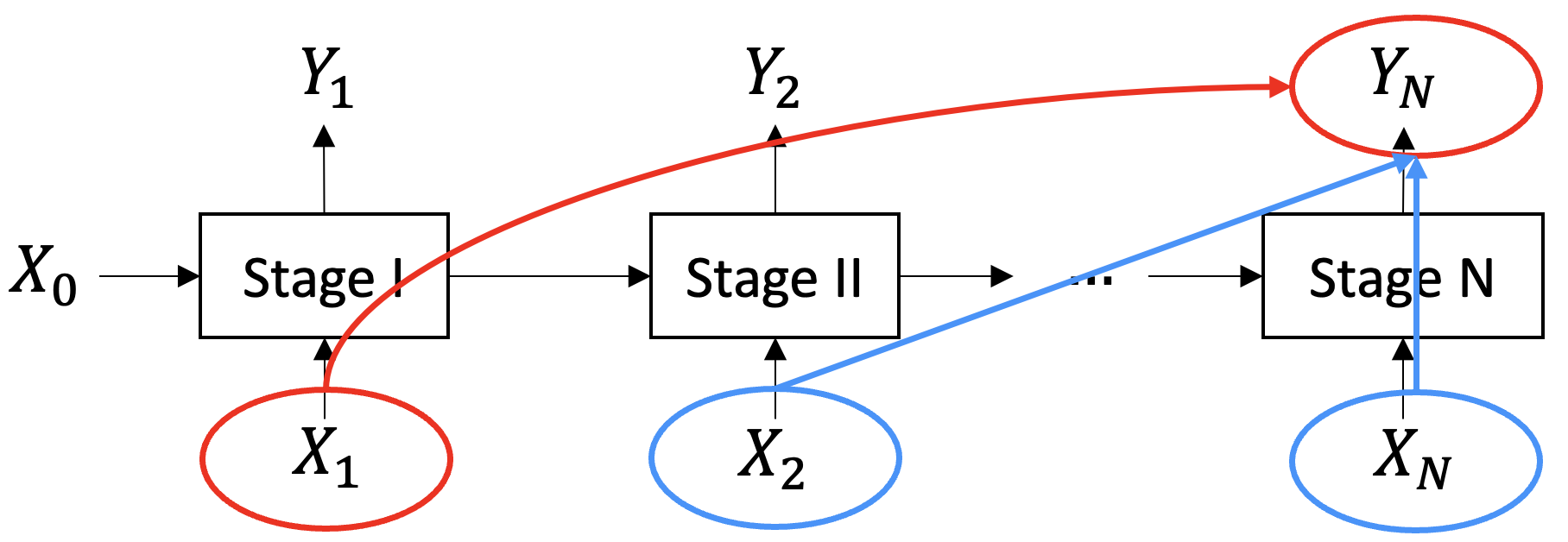}}
\caption{Feedforward quality control schemes anticipate the quality variation caused by disturbances in $X_1$ (red arrow) and plan for adjustments in $X_2$ through $X_N$ (blue arrows).}
\label{fig:fig2}
\end{figure}

In current MMSs, feedback control is commonly applied to enhance quality control. This approach involves monitoring quality variations and then manipulating process variables to mitigate future variations. While useful in many applications, its key drawback is the inability to provide timely corrections for MMSs, due to its dependency on the availability of quality measurements. On the other hand, feedforward control uses a prediction model to anticipate and counteract potential variations before they occur, making it a more proactive and sustainable approach. This study focuses on developing a feedforward control scheme to harness the capabilities of the SDK model. Fig. \ref{fig:fig2} illustrates this idea.

Without loss of generality, we assume that a product has already undergone processing operations from $1$ to $k$, and the goal of designing a feedforward controller is to optimize the downstream settings in operations $k+1$ to $N$ to minimize variations in product quality.

Given the SDK in Section \ref{sec: SDK}, the following equations can be used to predict product quality in MMSs:
\begin{equation*}
    \begin{aligned}
        H_0 &= \phi_0(X_0) 
        \quad &(\text{Initialization}) \\
        H_k &= f_k(H_{k-1}, X_k) 
        \quad &(\text{Propagation}) \\
        \tilde{Y}_k &= \text{MLP}_k(H_k) \quad &(\text{Prediction})
    \end{aligned}
\end{equation*}
where $f_k$ combines the encoding module and transition module. Note that each $H_k$ is a Gaussian random variable parameterized by $\epsilon_k$ as described earlier. Consider the situation where we need to compensate for a process fluctuation $\delta X_k$, whose impact on the local product quality is:
\begin{equation}
    \begin{aligned}
        H_k &= f_k(H_{k-1}, X_{\text{nom},k} + \delta X_k) \\
        \Delta \tilde{Y}_k &= \text{MLP}_k(H_k) - \text{MLP}_k(H_{\text{nom}, k})
    \end{aligned}
\end{equation}
where $X_{\text{nom},k}$ and $H_{\text{nom}, k}$ are respectively the process measurement and the quality indicator under the nominal operation. Given the information up to stage $S_k$, the impact of $\delta X_k$ on the downstream quality follows a recursive fashion:
\begin{equation} \label{quality_propagation_till_final}
    \begin{aligned}
        H_{k+1} & = f_{k+1}(H_k, X_{\text{nom}, k+1}) \\
        & \quad\quad\quad\quad \smash{\raisebox{-.5ex}{\vdots}}\\
        H_l & = f_l\big(f_{l-1} \ldots f_{k+1}(H_k, X_{\text{nom}, k+1}), \ldots ,X_{\text{nom},l}\big) \\
        \Delta \tilde{Y}_l &= \text{MLP}_l(H_{l}) - \text{MLP}_l(H_{\text{nom}, l})
    \end{aligned}
\end{equation}
where $l=k+1, \ldots, N$. Eq. \eqref{quality_propagation_till_final} is based on the assumption that the downstream operations beyond $S_k$ will still follow the nominal settings. Also, Eq. \eqref{quality_propagation_till_final} can be evaluated repeatedly at subsequent stages as the process continues, allowing for the consideration of multiple disturbances being introduced to the product quality. 

To formulate the feedforward control problem, we denote ${\Delta_Y = \big[ \Delta \tilde{Y}_{k+1}^\top, \ldots, \Delta \tilde{Y}_N^\top \big]^\top}$ to represent the aggregation of estimated downstream quality variations. To compensate for the variation caused by $\delta X_k$, one seeks a sequence of commands ${\Delta_X = \big[ \Delta X_{k+1}^\top, \ldots, \Delta X_N^\top \big]^\top}$ to adjust the downstream process parameters. Note that $\Delta_Y$ is a random vector induced from Gaussian variables ($H_k$), therefore the control objective is to minimize the expectation of variations. The feedforward control commands are obtained by solving the following optimization problem \footnote{One should notice the changes in the bold terms compared to the ACC version of this paper. The $\Delta X$ terms represent the adjustments needed for the downstream process parameters to compensate for the assessed quality variations.}:
\begin{equation} \label{quality_control_formulation}
    \begin{aligned}
        \min_{\Delta_X} \quad &\mathbb{E}(\Delta_Y^\top Q \Delta_Y) + \Delta_X^\top R \Delta_X 
        \\
        \text{s.t.} \quad & H_l  = f_l\big(f_{l-1} \ldots f_{k+1}(H_k, X_{\text{nom}, k+1} + \boldsymbol{\Delta X_{k+1}} ), \ldots ,X_{\text{nom},l} + \boldsymbol{\Delta X_l} \big)
        \\
        & \Delta \tilde{Y}_l = \text{MLP}_l(H_{l}) - \text{MLP}_l(H_{\text{nom}, l}) 
        \\
        &\Delta X_l \in \mathcal{X}_l.
    \end{aligned}
\end{equation}
$Q$ and $R$ are positive semi-definite weighting matrices, and $\mathcal{X}_l$ is the admissible region for the adjustments at each stage. The first term in the objective function minimizes the quality variations, and the second term avoids the system being overcontrolled. In order to perform fast estimates of $\mathbb{E}(\Delta_Y^\top Q \Delta_Y)$ to achieve speeds appropriate for online control, we approximate it numerically by setting ${\epsilon_k=0}$ for every stage. Eq. \eqref{quality_control_formulation} presents a nonlinear programming problem, which can be solved by using interior point methods; we use IPOPT by CasADi \cite{Andersson2019} in this study.

Formulation \eqref{quality_control_formulation} builds upon the same idea from \cite{izquierdo2007feedforward, yue2018surrogate}, but now extends to nonlinear MMSs using the SDK model in this work. Our formulation is also similar to model predictive control, albeit with a different focus. Instead of using models to propagate states over time, we use the SDK model to propagate quality information across stages in MMSs. When implementing the algorithm in real production, the optimization problem needs to be solved repeatedly as the product reaches each new stage. This is necessary to dynamically refine the adjustments based on the newly obtained process and quality information. The receding horizon for solving Eq. \eqref{quality_control_formulation} can be regarded as $(N-k)$ in our setting. However, this horizon can be adjusted to fixed values as required to meet specific needs. Since the control commands $\Delta_X$ modify the process parameters directly, the proposed control scheme essentially acts as a supervisory controller. The control commands are then communicated to the lower-level controllers and actuators, which will execute the adjustments.

\section{Case Study}
\label{sec: Case_study}
We demonstrate two case studies. The first is used to validate the SDK modeling method using a dataset obtained from a multistage R2R production line. The second focuses on validating the SDK feedforward control through a simulation.

\subsection{Validation of the Stochastic Deep Koopman Model}
\begin{figure}[ht]
\centerline{\includegraphics[width=\columnwidth]{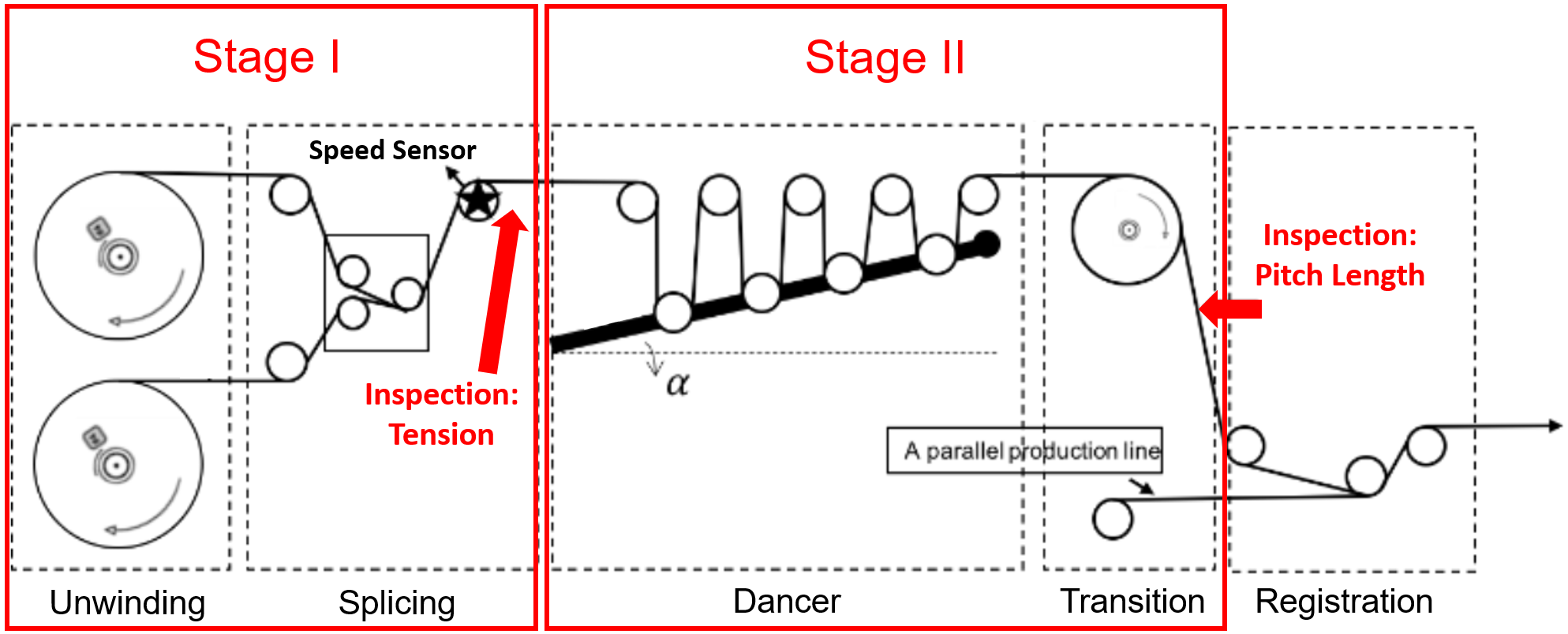}}
\caption{Layout of the R2R testbed \cite{shui2018twofold}.}
\label{fig:fig3}
\end{figure}

Initially introduced in \cite{shui2018twofold}, the dataset used in this case study is derived from a web handling process in a real-life R2R manufacturing system. To ensure confidentiality, all results have been normalized. The process configuration is depicted in Fig. \ref{fig:fig3}.

The R2R system is composed of five distinct operations: 1) unwinding, 2) splicing, 3) dancer, 4) transition, and 5) registration. The process involves releasing substrates from the unwind section, performing various intermediate processes, and stacking substrates into a multi-layer product at operation 5. Maintaining good product quality requires monitoring the pitch length of each substrate span at operation 5. A closed-loop control system utilizes this monitoring data to regulate substrate tension and roller speeds, ensuring consistent pitch lengths across all product spans. To facilitate proactive control actions, it is desirable to establish a model to estimate the quality metrics of each substrate.

The two quality characteristics of interest include: 1) substrate tension after splicing $t_1$, and 2) pitch length before registration $l_2$. The operations are grouped into two stages to enable efficient modeling of the R2R system, as shown in Fig. \ref{fig:fig3}. Our modeling task involves predicting the quality metrics using process measurements. Specifically, 16 process measurements are collected from stage I for tension prediction, and an additional 11 process measurements are gathered from stage II for pitch length prediction. 

The dataset used in this study is collected from 7 trials of operations. To prepare the sensor signals for modeling, a wavelet filter is applied to remove noise from the signals. The dataset is then divided into a training set (4 trials) and a testing set (3 trials). The training set is randomly shuffled, and 10\% of the shuffled data is separated as a validation set for monitoring the model's performance during training. The architecture of the stochastic Koopman model is selected based on its performance on the validation set. In this implementation, the latent spaces in the two stages have the same dimension, denoted as $d_{h,1}=d_{h,2}=40$. This choice is sufficiently large for the Koopman operators to discover a linear embedding of the quality propagation, which is reflected by a small prediction error. Loss weights are assigned as $\theta_1=\theta_2=0.01$, $\omega_1=\omega_2=5\times10^{-7}$, and $\rho_1=1$ and $\rho_2=10$. The algorithm is implemented using the Pytorch library, and trained using the stochastic gradient descent (SGD) optimizer on an Apple M1 chip laptop.

We demonstrate the performance of the SDK model through a comparison study with other  regression models:
\begin{enumerate}
\item Hybrid: a hybrid modeling approach from \cite{shui2018twofold} that includes a physical model component. Since this model incorporates the highest physics fidelity, we use it as a reference for all other results.
\item Artificial neural network (ANN):  a two-layer fully connected feedforward neural network with 64 hidden units and rectified linear unit (ReLU) activation for the hidden layer.
\item Random forest (RF): a model consisting of an ensemble of decision trees with hyperparameters set to: number of estimators $=30$, maximum depth $=10$, minimum samples at a leaf (fractional) $=0.01$. 
\item SDK: our model, introduced in Section \ref{sec: SDK}.
\end{enumerate}

\begin{figure}[ht]
\centerline{\includegraphics[width=\columnwidth]{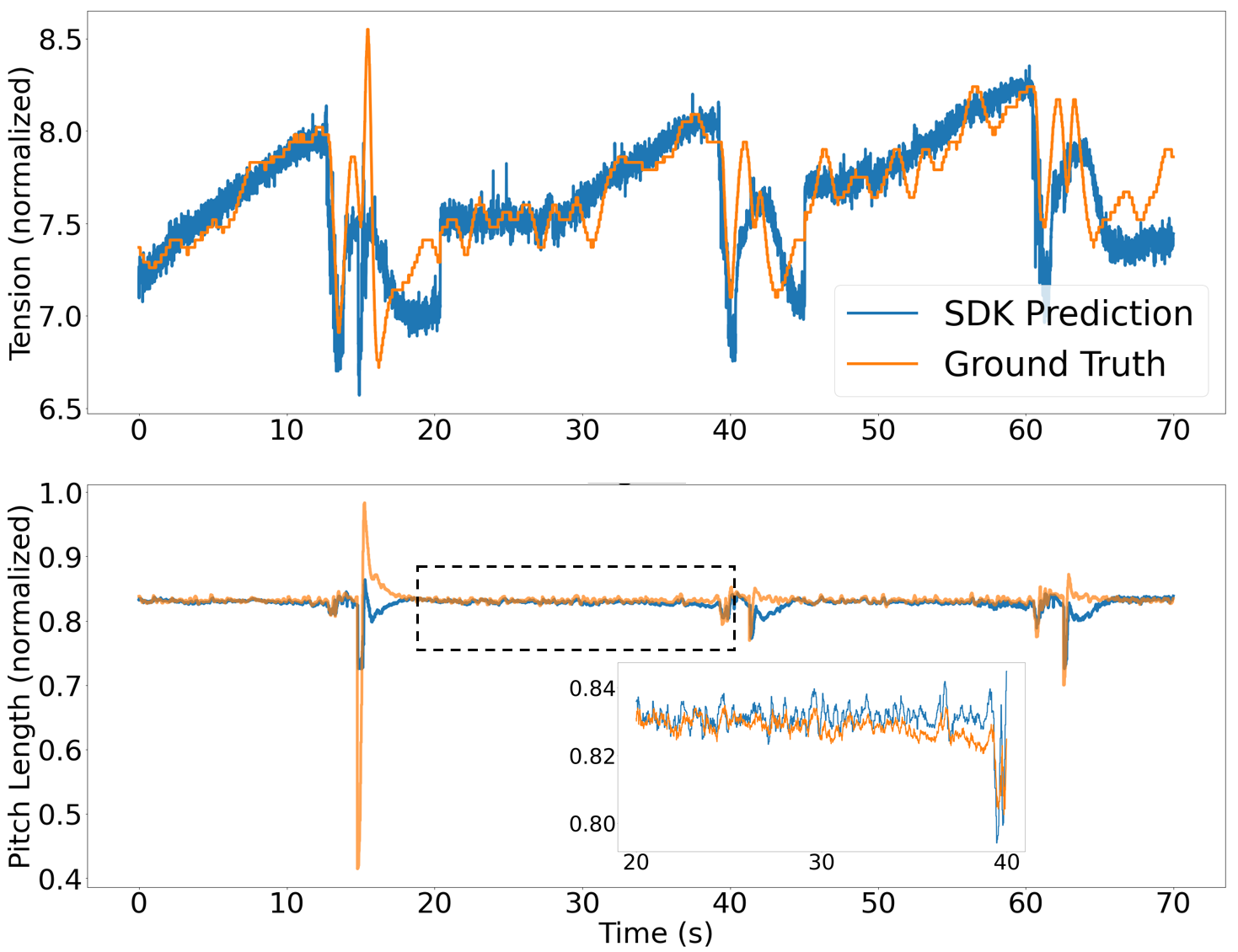}}
\caption{SDK predictions versus the ground truth, top: tension, and bottom: pitch length. A drift in pitch length is noticed in the enlarged section.}
\label{fig:fig4}
\end{figure}

Root mean squared errors (RMSE) on quality indices of the testing set are reported in Table \ref{table1} to illustrate the prediction performance of the above algorithms. The experiment on each data-driven method is repeated 10 times to obtain reproducible results, with the standard deviation shown as the $\pm$ value in the table. During the tests, ANN and RF use individual models to predict the quality measures from different stages, while the other models can obtain predictions under a unified framework. Prediction results on the testing set are shown in Fig. \ref{fig:fig4}.

\begin{table}
\centering
\setlength{\extrarowheight}{2pt}
\caption{Comparison of Prediction Errors}
\label{table1}
\setlength{\tabcolsep}{3pt}
\begin{tabular}{p{90pt}  p{125pt}  p{125pt} }
\toprule
\textbf{Model} & \textbf{Tension RMSE}    & \textbf{Pitch Length RMSE}  \\
\midrule
Hybrid \cite{shui2018twofold}      & \underline{0.2214}  & \underline{0.0102}\\
ANN                               & 0.2392 $\pm$ 0.0044       & 0.0251 $\pm$ 0.0021\\
RF                                & 0.2814 $\pm$ 0.0052       & 0.0240 $\pm$ 0.0002\\
\textbf{SDK}          & \textbf{0.2254 $\pm$ 0.0101}       & \textbf{0.0192 $\pm$ 0.0016}\\
\bottomrule
\end{tabular}
\end{table}

The hybrid model achieves the lowest prediction error for both stages, which aligns with our expectations as it includes a high-fidelity physics model. Our SDK model performs the best among the data-driven algorithms, reaching a similar level of accuracy in predicting tension compared to the hybrid method. However, it performs less effectively in predicting pitch length. In Fig. \ref{fig:fig4}, it is noticeable that pitch length prediction deviates further from the ground truth with time. This can be attributed to pitch length being influenced by the changing dynamics of R2R processes, making it challenging to predict without access to physics information. One promising future direction is thus to explore the integration of physical knowledge into the SDK model.

\subsection{Validation of the Quality Control Scheme}
\begin{figure}[ht]
\centerline{\includegraphics[width=\columnwidth]{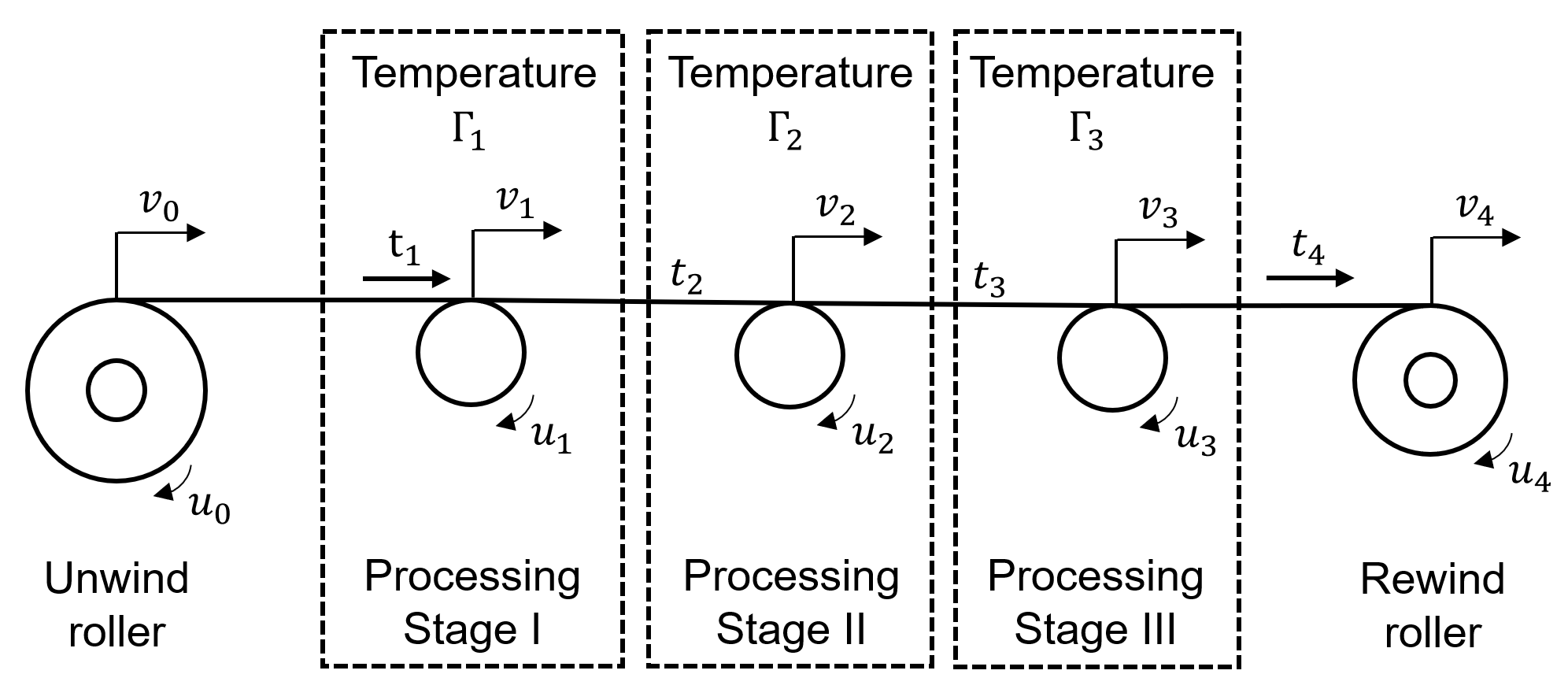}}
\caption{Schematic of the simulated R2R system.}
\label{fig:fig5}
\end{figure}

In this section, a simulation study is conducted using a R2R printing process. We focus on the three processing stages illustrated in Figure \ref{fig:fig5}.

The key parameters involved in this simulation include: 1) pitch lengths at the three stages $l_i$, 2) substrate tensions $t_i$, 3) roller speeds $v_i$, 4) speed deviations $\delta v_i$, 5) motor torques $u_i$, and 6) chamber temperatures $\Gamma_i$. The simulation is conducted based on the physical-induced tension and speed model described in \cite{chen2023robust}. In addition, the substrate's thermal dynamics are considered to account for pitch expansion under different temperature profiles. The pitch length dynamics is characterized by the following equation:
\begin{equation} \label{Koopman_quality_metric}
    l_i = \left(\frac{t_i}{AE}+1\right)l_0 + \alpha(\Gamma_i - \Gamma_0)l_0, \quad i=1,2,3
\end{equation}
where $l_0$ is the unstretched pitch length ($10 \mathrm{~cm}$ in our case), and $\Gamma_0$ is room temperature. $A$, $E$, $\alpha$ are the substrates' cross-sectional area, Young's modulus, and thermal expansion coefficient, respectively. The simulation is performed in MATLAB using the {\tt ode45} function. A total of 200 independent simulation trials are conducted, each lasting for $100 \mathrm{~s}$. In each trial, the system starts from a steady state and is then commanded to move to different operating conditions. A $\pm 5\%$ process noise is also added to the simulation. A robust tension controller from \cite{chen2023robust} is implemented to perform process control.

An SDK model aims to learn the stagewise quality propagation within the R2R system from the simulated data. Table \ref{table2} lists the input-output parameters fed to the SDK model.

\begin{table}[ht]
\centering
\setlength{\extrarowheight}{2pt}
\caption{Inputs and outputs of the SDK model.}
\label{table2}
\setlength{\tabcolsep}{3pt}
\begin{tabular}{p{90pt}  p{135pt}  p{75pt} }
\toprule
\textbf{Stage} & \textbf{Inputs}    & \textbf{Outputs}  \\
\midrule
Stage I         &$v_0,u_0,v_1, \delta v_1,u_1,\Gamma_1$     &$t_1,l_1$ \\
Stage II        &$v_2,\delta v_2,u_2,\Gamma_2$              &$t_2,l_2$ \\
Stage III       &$v_3,\delta v_3,u_3,\Gamma_3$              &$t_3,l_3$ \\
\bottomrule
\end{tabular}
\end{table}

\begin{figure}[ht]
\centerline{\includegraphics[width=0.9\columnwidth]{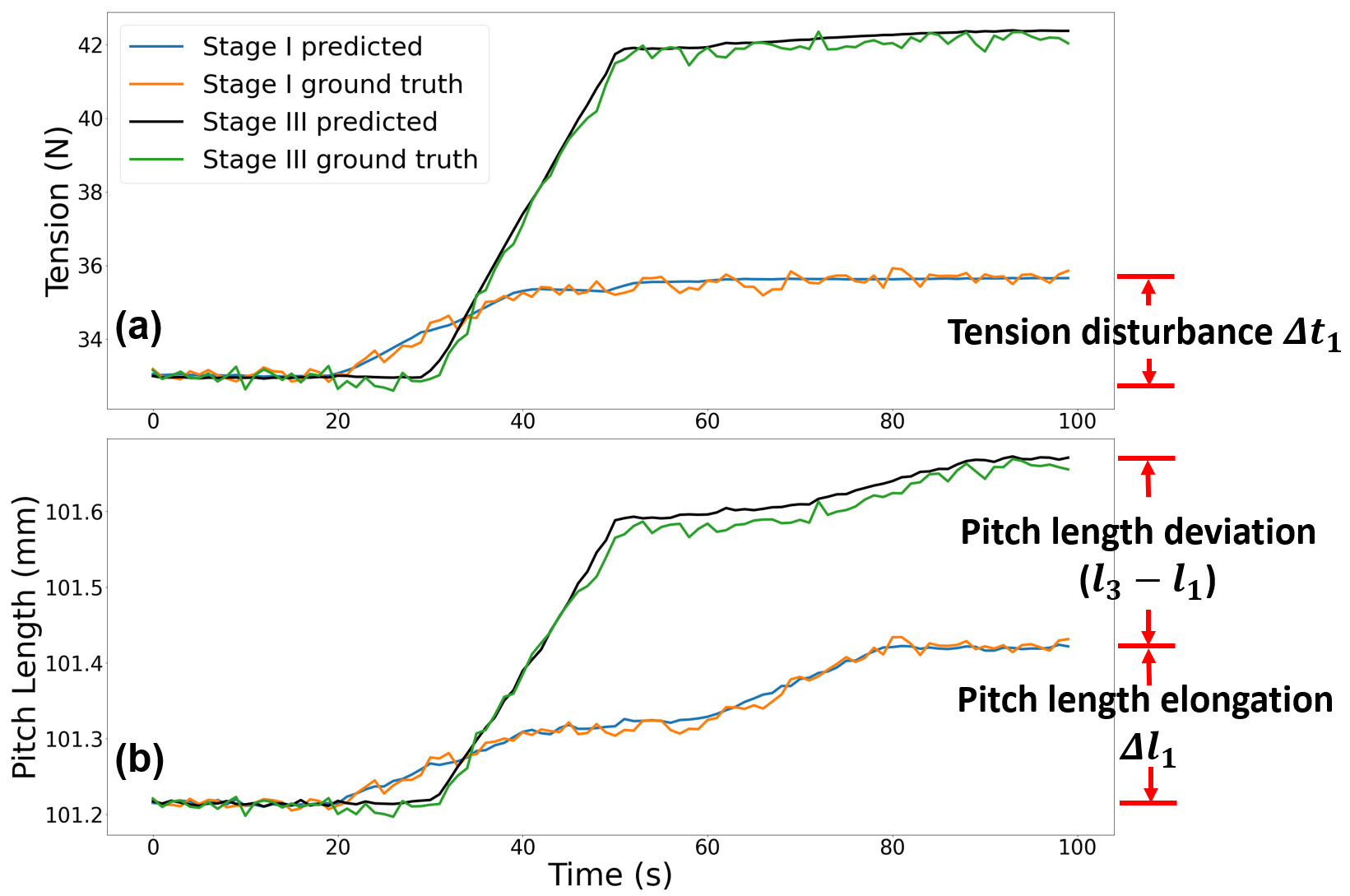}}
\caption{The SDK model provides accurate predictions. (a) Different tensions are applied to stages I and III; (b) Stages I and III end up having different pitch lengths.}
\label{fig:fig6}
\end{figure}

We expect the model to learn how the process parameters affect the tensions and pitch lengths without relying on the underlying physics. The SDK model achieves accurate predictions as illustrated in Fig. \ref{fig:fig6}. As implied by Eq. \eqref{Koopman_quality_metric}, the temperature and tension changes will both contribute to elongations in pitch length. When various temperatures and tensions are applied to different stages, a deviation between pitch lengths at stages I and III is observed, denoted by $(l_3-l_1)$ in Fig. \ref{fig:fig6}(b). This deviation in pitch lengths will cause pattern distortions in R2R printing processes.

\begin{figure}[ht]
\centerline{\includegraphics[width=0.75\columnwidth]{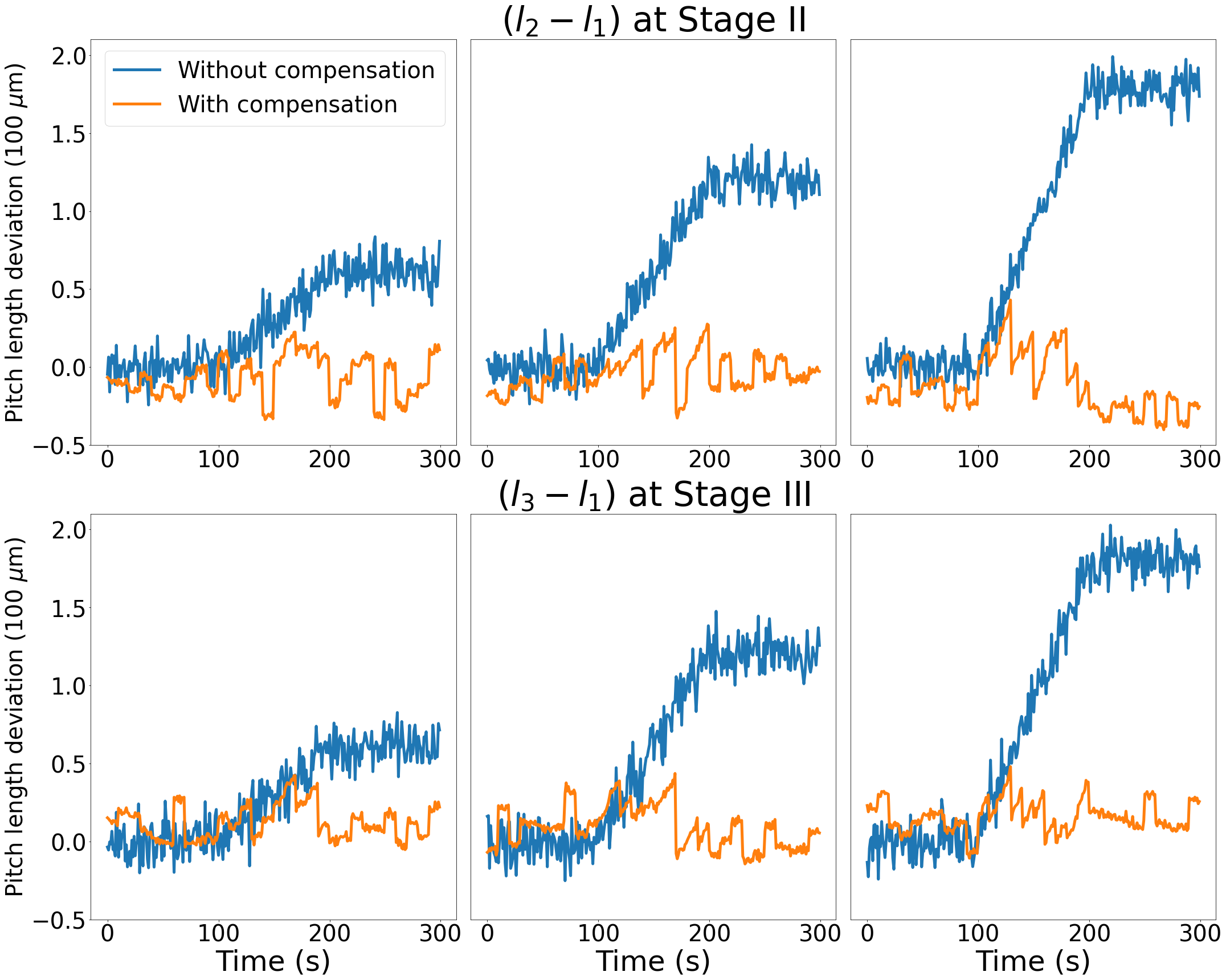}}
\caption{The disturbance in $\Gamma_1$ causes pitch length deviations in stages II and III. The feedforward controller compensates for this quality variation.}
\label{fig:fig7}
\end{figure}

The control objective is that, for each substrate span, its pitch length at different stages should be synchronized with respect to $l_1$, meaning that deviations $(l_2-l_1)$ and $(l_3-l_1)$ should be minimized. Three new simulation trials are conducted to demonstrate the system performance after the feedforward controller is deployed. Each testing trial lasts for $300 \mathrm{~s}$. During the test, disturbances in the temperature profile $\Gamma_1$ are commanded. The temperature variations result in pitch length deviations across different stages, as depicted by the blue curves in Fig. \ref{fig:fig7}. The remaining process parameters are considered controllable to compensate for these quality variations. Specifically, the controller periodically evaluates the quality metrics and suggests adjusting the downstream process settings every $10 \mathrm{~s}$. Computations to solve \eqref{quality_control_formulation} for each control cycle take less than $10 \mathrm{~ms}$ on a laptop.

\begin{figure}
\centerline{\includegraphics[width=0.75\columnwidth]{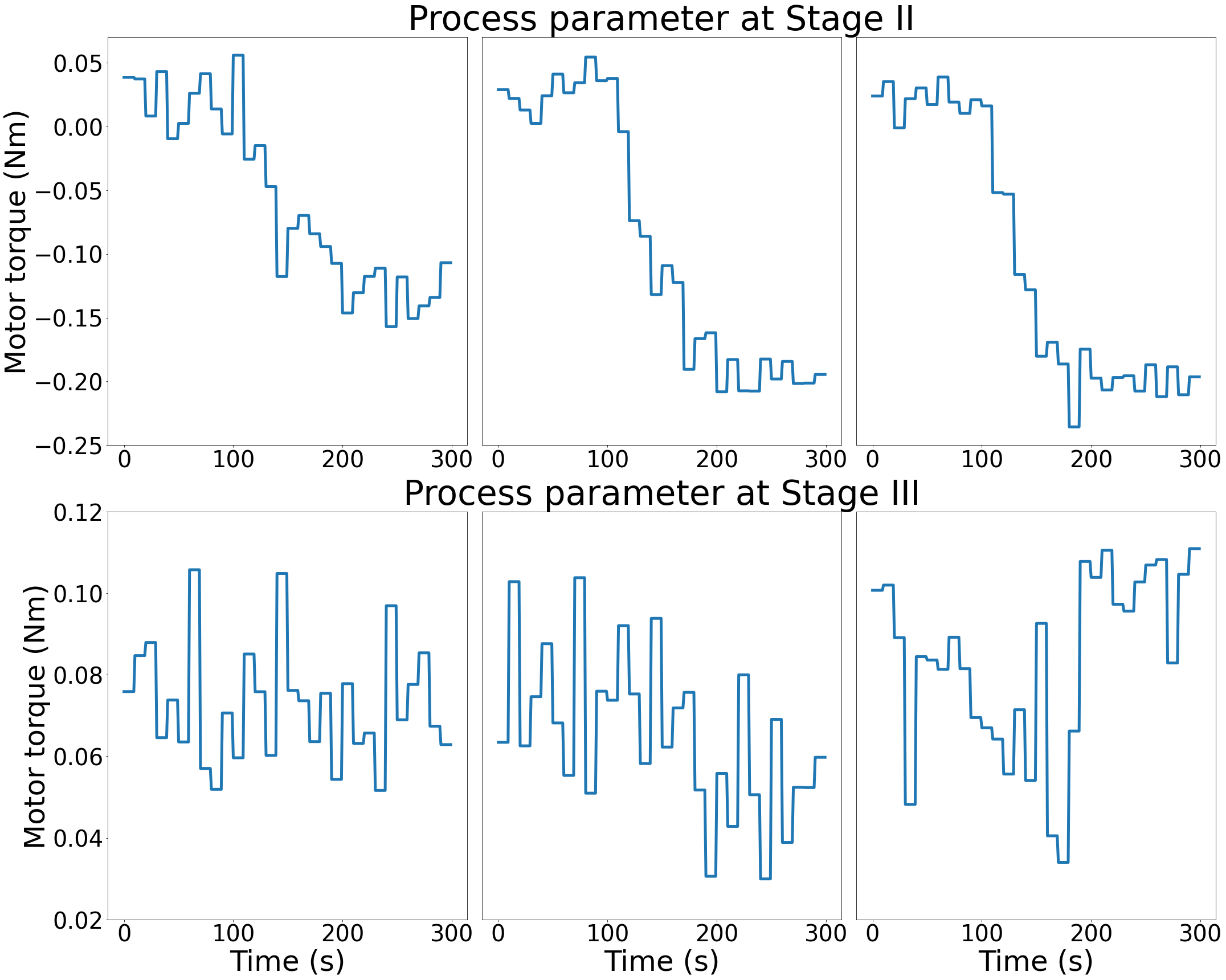}}
\caption{The motor torques, as the key process parameters, are dynamically adjusted by the controller.}
\label{fig:fig8}
\end{figure}

Figure \ref{fig:fig7} shows that when not applying the feedforward controller, disturbances can result in pitch length variations of $60$, $120$, and $180\mathrm{\mu m}$ for both stages II and III in the three test trials. However, with the controller, this trend can be predicted and eliminated. Specifically, the deviations between pitch lengths are regulated within $40\mathrm{\mu m}$ when different levels of temperature disturbances are present. Furthermore, Figure \ref{fig:fig8} demonstrates the collaborative behavior of the motors at stages II and III in response to disturbances. The feedforward controller supervises the process controllers and actuators to work in cooperation. Set points of process parameters are dynamically adjusted to counteract the effects of temperature changes and maintain the desired pitch lengths. This showcases the capability of the proposed feedforward controller to optimize the overall performance of the production line by considering the interdependencies between different stages. In this case, the interdependencies among stages are learned directly from the data and captured in the Koopman transition modules, minimizing the reliance on human knowledge.

\section{Conclusion}
\label{sec: Conclusion}
This paper presents 
a novel feedforward controller for minimizing quality variations in MMSs, leveraging an SDK predictive model. Two R2R case studies are presented to validate the effectiveness of the proposed method and highlight its potential for improving the operation of R2R production lines. The key contributions and findings are as follows.
\begin{enumerate}
\item The SDK framework achieves high accuracy in predicting stage-by-stage product quality in MMSs.
\item The feedforward quality control scheme can effectively minimize quality variations by adjusting process parameters in real time.
\item The modeling-control synthesis can be applied to nonlinear MMSs, requiring minimal physical information.
\end{enumerate}

The proposed method exhibits several limitations. Firstly, the synthesis needs a large amount of data for training, posing challenges in data collection for industrial applications. Secondly, the control algorithm is designed based on the assumption that the downstream operations can compensate for upstream variations. This stagewise compensability needs further study. 
Furthermore, enhancing the quality control scheme to better handle process noise and model uncertainty will be essential to improve its robustness and reliability.

\bibliography{reference}

\begin{thebibliography}{10}

\bibitem{abellan2012quality}
J.~V. Abellan-Nebot, J.~Liu, and F.~R. Subir{\'o}n.
\newblock Quality prediction and compensation in multi-station machining processes using sensor-based fixtures.
\newblock {\em Robotics and Computer-Integrated Manufacturing}, 28(2):208--219, 2012.

\bibitem{Andersson2019}
J.~A.~E. Andersson, J.~Gillis, G.~Horn, J.~B. Rawlings, and M.~Diehl.
\newblock {CasADi} -- {A} software framework for nonlinear optimization and optimal control.
\newblock {\em Mathematical Programming Computation}, 11(1):1--36, 2019.

\bibitem{balakrishnan2021stochastic}
K.~Balakrishnan and D.~Upadhyay.
\newblock Stochastic adversarial koopman model for dynamical systems.
\newblock {\em arXiv preprint arXiv:2109.05095}, 2021.

\bibitem{chen2023stochastic}
Z.~Chen, H.~Maske, H.~Shui, D.~Upadhyay, M.~Hopka, J.~Cohen, X.~Lai, X.~Huan, and J.~Ni.
\newblock Stochastic deep koopman model for quality propagation analysis in multistage manufacturing systems.
\newblock {\em Journal of Manufacturing Systems}, 71:609--619, 2023.

\bibitem{chen2023robust}
Z.~Chen, B.~Qu, B.~Jiang, S.~R. Forrest, and J.~Ni.
\newblock Robust constrained tension control for high-precision roll-to-roll processes.
\newblock {\em ISA transactions}, 136:651--662, 2023.

\bibitem{djurdjanovic2017multistage}
D.~Djurdjanovi{\'c}, Y.~Jiao, and V.~Majstorovi{\'c}.
\newblock Multistage manufacturing process control robust to inaccurate knowledge about process noise.
\newblock {\em CIRP Annals}, 66(1):437--440, 2017.

\bibitem{djurdjanovic2007online}
D.~Djurdjanovi{\'c} and J.~Ni.
\newblock Online stochastic control of dimensional quality in multistation manufacturing systems.
\newblock {\em Proceedings of the Institution of Mechanical Engineers, Part B: Journal of Engineering Manufacture}, 221(5):865--880, 2007.

\bibitem{dreyfus2022virtual}
P.~Dreyfus, F.~Psarommatis, G.~May, and D.~Kiritsis.
\newblock Virtual metrology as an approach for product quality estimation in industry 4.0: a systematic review and integrative conceptual framework.
\newblock {\em International Journal of Production Research}, 60(2):742--765, 2022.

\bibitem{izquierdo2007feedforward}
L.~E. Izquierdo, J.~Shi, S.~J. Hu, and C.~W. Wampler.
\newblock Feedforward control of multistage assembly processes using programmable tooling.
\newblock {\em Trans. NAMRI/SME}, 35:295--302, 2007.

\bibitem{kingma2013auto}
D.~P. Kingma and M.~Welling.
\newblock Auto-encoding variational bayes.
\newblock {\em arXiv preprint arXiv:1312.6114}, 2013.

\bibitem{lee2022stream}
J.~Lee, P.~Gore, X.~Jia, S.~Siahpour, P.~Kundu, and K.~Sun.
\newblock Stream-of-quality methodology for industrial internet-based manufacturing system.
\newblock {\em Manufacturing Letters}, 34:58--61, 2022.

\bibitem{lusch2018deep}
B.~Lusch, J.~N. Kutz, and S.~L. Brunton.
\newblock Deep learning for universal linear embeddings of nonlinear dynamics.
\newblock {\em Nature Communications}, 9(1):1--10, 2018.

\bibitem{peres2019multistage}
R.~S. Peres, J.~Barata, P.~Leitao, and G.~Garcia.
\newblock Multistage quality control using machine learning in the automotive industry.
\newblock {\em IEEE Access}, 7:79908--79916, 2019.

\bibitem{shi2023process}
J.~Shi.
\newblock In-process quality improvement: Concepts, methodologies, and applications.
\newblock {\em IISE transactions}, 55(1):2--21, 2023.

\bibitem{shui2018twofold}
H.~Shui, X.~Jin, and J.~Ni.
\newblock Twofold variation propagation modeling and analysis for roll-to-roll manufacturing systems.
\newblock {\em IEEE Transactions on Automation Science and Engineering}, 16(2):599--612, 2018.

\bibitem{wang2023production}
P.~Wang, H.~Qu, Q.~Zhang, X.~Xu, and S.~Yang.
\newblock Production quality prediction of multistage manufacturing systems using multi-task joint deep learning.
\newblock {\em Journal of Manufacturing Systems}, 70:48--68, 2023.

\bibitem{yan2021deep}
H.~Yan, N.~D. Sergin, W.~A. Brenneman, S.~J. Lange, and S.~Ba.
\newblock Deep multistage multi-task learning for quality prediction of multistage manufacturing systems.
\newblock {\em Journal of Quality Technology}, 53(5):526--544, 2021.

\bibitem{yue2018surrogate}
X.~Yue and J.~Shi.
\newblock Surrogate model--based optimal feed-forward control for dimensional-variation reduction in composite parts' assembly processes.
\newblock {\em Journal of Quality Technology}, 50(3):279--289, 2018.

\bibitem{zhang2021path}
D.~Zhang, Z.~Liu, W.~Jia, H.~Liu, and J.~Tan.
\newblock Path enhanced bidirectional graph attention network for quality prediction in multistage manufacturing process.
\newblock {\em IEEE Transactions on Industrial Informatics}, 18(2):1018--1027, 2021.

\bibitem{zhang2022contrastive}
D.~Zhang, Z.~Liu, W.~Jia, H.~Liu, and J.~Tan.
\newblock Contrastive decoder generator for few-shot learning in product quality prediction.
\newblock {\em IEEE Transactions on Industrial Informatics}, 2022.

\bibitem{zhao2023novel}
L.~Zhao, B.~Li, and Y.~Yao.
\newblock A novel predict-prevention quality control method of multi-stage manufacturing process towards zero defect manufacturing.
\newblock {\em Advances in Manufacturing}, pages 1--15, 2023.

\end{thebibliography}
\bibliographystyle{abbrv}

\end{document}